\documentclass[aps,pra,reprint,amsmath,amssymb,superscriptaddress,nofootinbib,longbibliography]{revtex4-2}

\usepackage{graphicx}

\usepackage[utf8]{inputenc}
\usepackage[T1]{fontenc}
\usepackage{xcolor}
\usepackage{soul}

\usepackage{amssymb}
\usepackage{bm}

\usepackage{enumitem}

\begin{document}

\title{Pressure-temperature magnetostructural phase diagrams of slowly cooled
Co$_{1-x}$Cu$_x$MnGe $(0.05 \leq x \leq 0.35)$}

\author{Ryszard Duraj}
\affiliation{Institute of Physics, Cracow University of Technology, Podchor\k{a}\.zych 1,
PL-30-084 Krak\'ow, Poland}
\author{Aleksandra Deptuch}
\affiliation{Institute of Nuclear Physics Polish Academy of
Sciences, Radzikowskiego 152, PL-31-342 Krak\'ow, Poland}
\author{Andrzej Szytu\l{}a}
\affiliation{M.~Smoluchowski Institute of Physics, Jagiellonian University,
prof. Stanis\l{}awa \L{}ojasiewicza 11, PL-30-348 Krak\'ow, Poland}
\author{Bogus\l{}aw Penc}
\affiliation{M.~Smoluchowski Institute of Physics, Jagiellonian University,
prof. Stanis\l{}awa \L{}ojasiewicza 11, PL-30-348 Krak\'ow, Poland}
\author{Stanis\l{}aw Baran}
\email{stanislaw.baran@uj.edu.pl}
\affiliation{M.~Smoluchowski Institute of Physics, Jagiellonian University,
prof. Stanis\l{}awa \L{}ojasiewicza 11, PL-30-348 Krak\'ow, Poland}

\date{\today}

\begin{abstract}

Polycrystalline samples of Co$_{1-x}$Cu$_x$MnGe ($x=0.05, 0.10, 0.15, 0.22$ and 0.35), prepared
by arc melting under argon atmosphere, have been annealed at 1123~K with final furnace
cooling. The samples have been investigated by powder X-ray diffraction (in function of temperature)
and ac magnetic measurements (in function of temperature and applied hydrostatic pressure up to
12~kbar). On the basis of the experimental data, the $(p,T)$ phase diagrams have been
determined. For the low Cu content ($x=0.05$, 0.10 and 0.15), the compounds show a martensitic
transition between the low-temperature orthorhombic crystal structure of the TiNiSi-type (space group:
$Pnma$) and the high-temperature hexagonal structure of the Ni$_2$In-type (space group: $P6_3/mmc$).
For the high Cu content ($x=0.22$ and 0.35) only the hexagonal structure is observed.
All compounds undergo a transition from para- to ferromagnetic state with decreasing temperature
(in case of $x=0.22$ through an intermediate antiferromagnetic phase). The para- to ferromagnetic
transition is fully coupled with the martensitic one for $x=0.05$ at the intermediate
pressure range (6~kbar $\le p \le 8$~kbar). Partial magnetostructural coupling is observed for $x=0.10$
at ambient pressure. The Curie temperature at ambient pressure decreases from 313~K for $x=0.05$ 
(in the orthorhombic phase) to about 250~K for the remaining compounds (in the hexagonal phase).
For the Co$_{0.85}$Cu$_{0.15}$MnGe compound, entropy change associated with the martensitic transition
has been calculated using Clausius-Clapeyron equation.

\bigskip

\noindent \textbf{keywords}: intermetallic compounds, X-ray diffraction (XRD), magnetic materials,
magnetic properties, pressure-temperature phase diagram

\end{abstract}

\maketitle

\section{Introduction}
\label{intro}

The magnetostructural transformation materials, i.e. the materials that undergo the
crystallographic and magnetic phase transitions simultaneously, are of considerable
attention not only for their importance in fundamental physics, but also for their
applications as multifunctional materials. Magnetostructural transformation, originating
from coupling between spins and lattice, is found in a number of ternary MM'Ge compounds,
where M and M' are 3d transition elements. CoMnGe and its derivative compounds form
a family of compounds whose magnetostructural properties can be easily tuned by
changing composition, thermodynamic parameters, sample preparation procedure, etc. The CoMnGe
parent compound crystallizes in two crystal structures: an orthorhombic one of the TiNiSi-type
(space group: $Pnma$, No. 62) and a hexagonal one of the Ni$_2$In-type (space group: $P6_3/mmc$,
No. 194), below and above the martensitic phase transition temperature ($T_s$) above 400~K,
respectively (the exact transition temperature depends on sample preparation procedure)~\cite{johnson1975diffusionless}.
The compound in both crystal variants is ferromagnetic with the Curie temperatures ($T_C$) of 337 and
283~K and magnetic moments per molecule equal to 3.71(2) and 2.80(5)~$\mu_B$
for the orthorhombic and hexagonal phases, respectively (see
Table~2 in \cite{KAPRZYK1990267}).
Neutron diffraction data indicate existence of a ferromagnetic order with the magnetic moment
in the Mn sublattice present in both phases, while the Co atoms possess magnetic moments
only in the orthorhombic phase~\cite{NIZIOL1982281,SZYTULA1981176}.
CoMnGe is located in vicinity of a critical point in the magnetostructural phase diagram for
pseudoternary germanides M$_{1-x}$M'$_x$MnGe (see Fig.~1 in Ref.~\cite{BECKMAN1991181}).
In the stoichiometric CoMnGe compound a magnetostructural coupling at ambient pressure is absent
as $T_s$ and $T_C$ are well separated. The magnetostructural transition can be induced by
application of external pressure exceeding 6~kbar~\cite{NIZIOL1983205}. Alternatively, the
magnetostructural coupling can be achieved by doping of substitutional and/or interstitial atoms
as well as introducing metal vacancies~\cite{ZENG2014101}.
Investigation of magnetic properties of CoMnGe and its derivatives is of great importance 
due to large magnetocaloric effect around room temperature~\cite{HAMER20093535,Trung_App_Phys_Lett_96,Trung_App_Phys_Lett_96_172504}.

Recently, it has been reported that doping Cu atoms into the Co sublattice in Co$_{1-x}$Cu$_x$MnGe
$(0 \leq x \leq 0.5)$ leads to complex magnetic behavior including appearance of new magnetic phases
and magnetostructural coupling~\cite{ZHANG2017531,PAL201922}. These intriguing properties have inspired us
to undertake the current study in which we report investigation of physical properties of
Co$_{1-x}$Cu$_x$MnGe (0.05 $\leq x \leq 0.35)$ by X-ray diffraction and ac magnetic susceptibility
measurements under hydrostatic pressure. On the basis of these data, the magnetostructural phase diagrams
$(p,T)$ are determined. This work is a continuation of our broader scientific
project concentrated on the role of external pressure on physical properties of the pseudoternary
M$_{1-x}$M'$_x$MnGe systems~\cite{DURAJ2018449,DEPTUCH2022114823}.

\section{Experimental details and results}

\subsection{Crystal structure}

Polycrystalline samples of Co$_{1-x}$Cu$_x$MnGe ($x=0.05, 0.10, 0.15, 0.22$ and 0.35) have been prepared
by arc melting of constituent elements (purity better than 99.9 wt \%) under argon atmosphere, followed by
annealing in high vacuum ($\sim$ $10^{-3}$ mbar) for 5 days at 1123~K (850~$^\circ$C) and subsequent furnace
cooling down to room temperature. Sample quality has been tested at room temperature by powder X-ray
diffraction (XRD) using X'Pert PRO (PANalytical) diffractometer (CuK$_{\alpha}$ radiation).

In order to investigate temperature evolution of the XRD patterns of Co$_{1-x}$Cu$_x$MnGe
(0.05 $\leq x \leq 0.35)$, the powder samples have been placed in capillaries mounted in Empyrean~2 (PANalytical) diffractometer (CuK$\alpha$
radiation, parallel incident beam, geometry of a horizontal rotating capillary,
$2\theta =$ 15--95$^\circ$ or 28--90$^\circ$). Temperature has been controlled with the Cryostream 700 Plus
(Oxford Cryosystems) attachment with 5~min. of stabilization before collection of each pattern. The patterns
have been collected on heating and subsequent cooling between 80 and 480~K. 

\begin{figure}[!ht]
\begin{center}
\includegraphics[bb=14 14 581 865, width=\columnwidth]
       {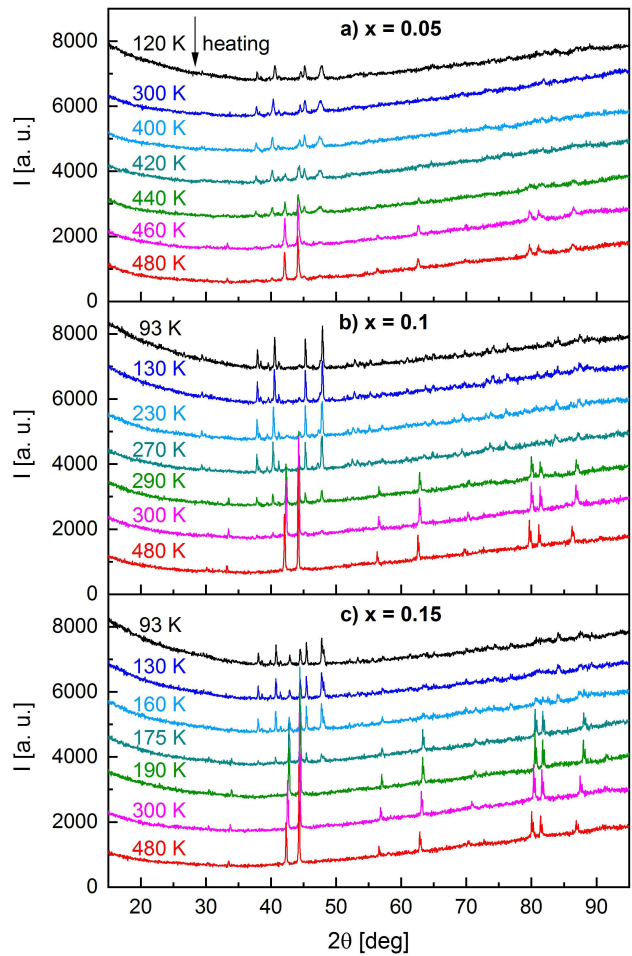}
\end{center}
\caption{\label{fig:XRD_sample_patterns}
Representative powder XRD patterns of Co$_{1-x}$Cu$_x$MnGe ($x =$ 0.05--0.15) collected on heating.}
\end{figure}

\begin{figure}[!ht]
\begin{center}
\includegraphics[bb=14 14 723 440, width=\columnwidth]
	{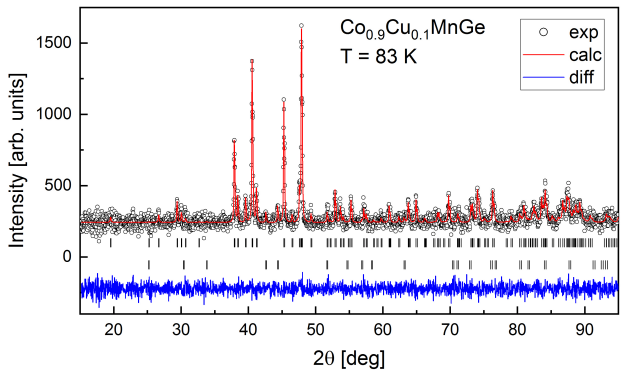}
\end{center}
\caption{\label{fig:XRD_pattern_x=0.10_83K}
XRD pattern of Co$_{0.9}$Cu$_{0.1}$MnGe collected at 83~K together with the results of Rietveld
refinement. For better clarity, the refined background has been subtracted from both the experimental
and calculated patterns after the final refinement. The first and second rows of the vertical bars indicate
the Bragg reflection positions originating from the low-temperature orthorhombic phase ($Pnma$, weight fraction
96.4(1.4)\%) and the high-temperature hexagonal phase ($P6_3/mmc$, weight fraction 3.6(5)\%), respectively.}
\end{figure}

Representative X-ray diffraction patterns (XRD) of Co$_{1-x}$Cu$_x$MnGe ($x =$ 0.05, 0.10 and 0.15)
collected at selected temperatures are shown in Fig.~\ref{fig:XRD_sample_patterns}. For all compounds a clear
change of the crystal structure with increasing temperature is evident. In order to follow temperature
evolution of the crystal structure, a Rietveld-type computer program FullProf~\cite{RODRIGUEZCARVAJAL199355}
has been used to process the XRD data. Fig.~\ref{fig:XRD_pattern_x=0.10_83K} shows a sample XRD pattern
together with the results of Rietveld refinement. The compounds investigated in this work crystallize
in two types of the crystal structure:

\begin{itemize}

\item the low-temperature orthorhombic phase of the TiNiSi-type (space group $Pnma$) in which each element occupies
the $4c$ site $(x,\frac{1}{4},z)$ with different values of the $x$ and $z$ parameters,

\item the high-temperature hexagonal phase of the Ni$_2$In-type (space group $P6_3/mmc$) with the Co atoms occupying
the $2d$ site $(\frac{1}{3},\frac{2}{3},\frac{3}{4})$, the Mn atoms at the $2a$ site $(0,0,0)$ and the Ge atoms at
the $2c$ site $(\frac{1}{3},\frac{2}{3},\frac{1}{4})$.

\end{itemize}

\begin{figure*}[!ht]
\begin{center}
\includegraphics[bb=14 14 1149 865, width=2\columnwidth]
	{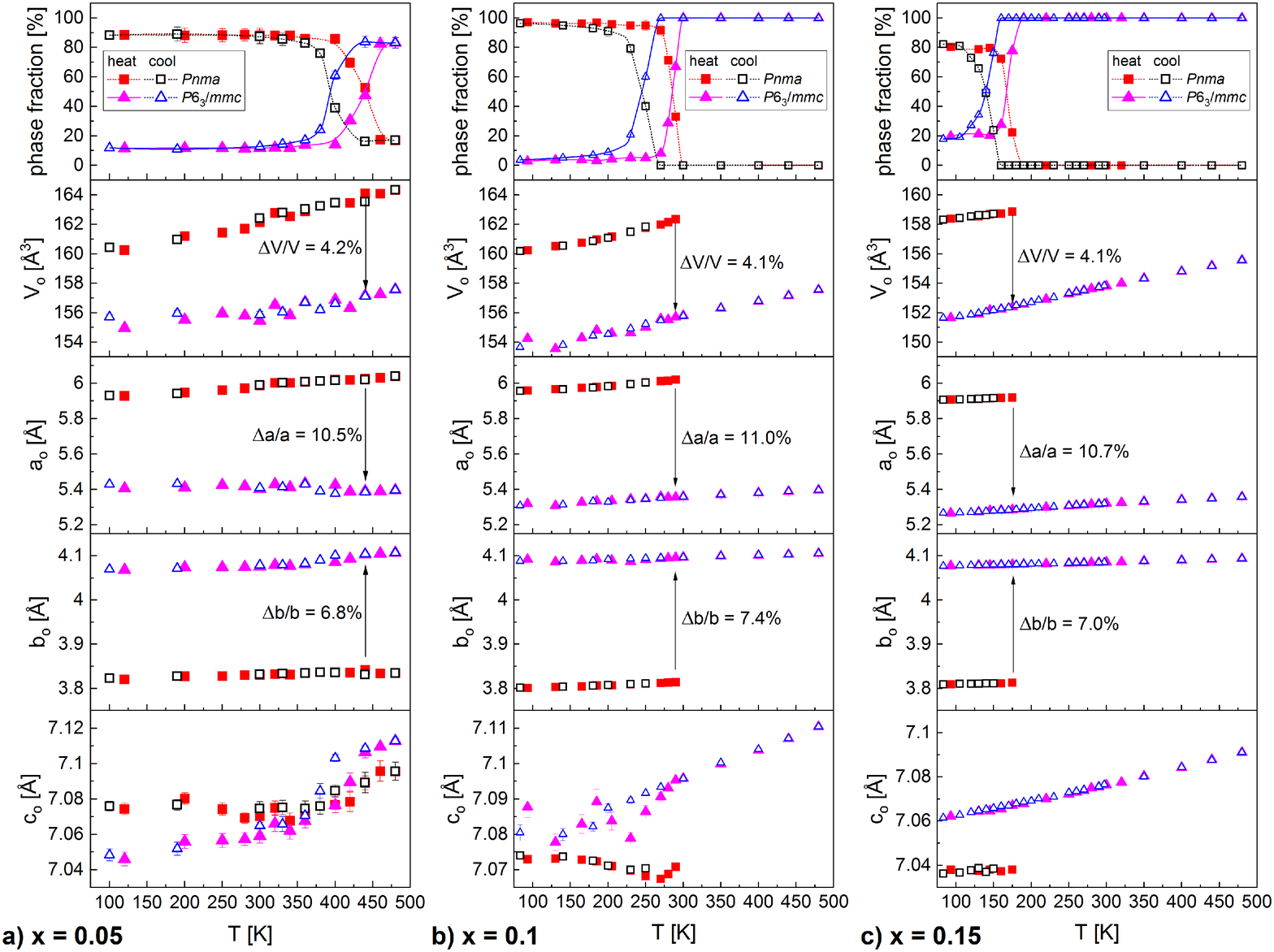}
\end{center}
\caption{\label{fig:temp_depend_of_cryst_param_x=0.05-0.15}
Thermal evolution of the phase fraction (weight percentage), volume of the unit cell and lattice constants
for Co$_{1-x}$Cu$_x$MnGe ($x =$ 0.05--0.15). Unit cell parameters of the hexagonal phase have been
recalculated to those of the orthorhombic phase using the following relations: $a_o=c_h$, $b_o=a_h$,
$c_o=\sqrt{3}a_h$ and $V_o=2V_h$.}
\end{figure*}

\begin{table*}
\caption{\label{tab:latticeort}
Lattice parameters $a_o$, $b_o$ and $c_o$, unit cell volume $V_o$, weight fraction and agreement factors of Rietveld
refinement obtained for the orthorhombic phase of Co$_{1-x}$Cu$_x$MnGe from the X-ray diffraction data collected
at the lowest investigated temperature and close to the martensitic transition for $x$ = 0.05-0.15 on heating.
}
\begin{footnotesize}
\begin{tabular*}{\textwidth}{@{\extracolsep{\fill}}ccccccccc}
$x$ & $T$ [K] & $a_o$ [\AA{}] & $b_o$ [\AA{}] & $c_o$ [\AA{}] & $V_o$ [\AA{}$^3$] & fraction [\%] & $R_{Bragg}$ [\%] & $R_f$ [\%]\\ \hline
0.05 & 100 & 5.931(3) & 3.823(2) & 7.076(3) & 160.4(2) & 88.2(2.3) & 8.7 & 6.8\\
 & 440 & 6.026(5) & 3.842(3) & 7.089(4) & 164.1(2) & 52.6(3.1) & 18.3 & 15.6\\
0.1 & 83 & 5.9569(7) & 3.8013(4) & 7.0740(8) & 160.18(3) & 96.4(1.4) & 9.4 & 8.6\\
 & 290 & 6.0204(9) & 3.8139(4) & 7.0708(9) & 162.35(4) & 33.0(1.3) & 20.2 & 20.8\\
0.15 & 83 & 5.9069(9) & 3.8091(4) & 7.0363(8) & 158.32(4) & 82.2(2.0) & 12.3 & 11.0\\
 & 175 & 5.919(2) & 3.8130(5) & 7.038(1) & 158.85(5) & 22.3(1.1) & 36.0 & 34.7\\
\end{tabular*}
\end{footnotesize}
\end{table*}

\begin{table*}
\caption{\label{tab:latticehex}
Lattice parameters $a_h$ and $c_h$, unit cell volume $V_h$, weight fraction and agreement factors of Rietveld refinement obtained for the 
hexagonal phase of Co$_{1-x}$Cu$_x$MnGe from the X-ray diffraction data collected at 480 K for $x$ = 0.05-0.35 and close to the martensitic 
transition for $x$ = 0.05-0.15 on heating.
}
\begin{footnotesize}
\begin{tabular*}{\textwidth}{@{\extracolsep{\fill}}cccccccc}
$x$ & $T$ [K] & $a_h$ [\AA{}] & $c_h$ [\AA{}] & $V_h$ [\AA{}$^3$] & fraction [\%] & $R_{Bragg}$ [\%] & $R_f$ [\%]\\ \hline
0.05 & 440 & 4.103(2) & 5.390(3) & 78.59(7) & 47.4(2.0) & 6.9 & 8.5\\
 & 480 & 4.1065(9) & 5.394(2) & 78.78(4) & 100 & 6.6 & 10.5\\
0.1 & 290 & 4.0962(2) & 5.3569(3) & 77.852(8) & 67.0(1.6) & 6.7 & 8.3\\
 & 480 & 4.1052(2) & 5.3976(3) & 78.778(8) & 100 & 7.9 & 9.7\\
0.15 & 175 & 4.0802(2) & 5.2855(2) & 76.204(5) & 77.7(1.8) & 7.9 & 10.1\\
 & 480 & 4.0940(2) & 5.3582(3) & 77.777(8) & 100 & 8.6 & 11.7\\
0.22 & 480 & 4.1115(2) & 5.4070(3) & 79.156(8) & 100 & 7.1 & 11.7\\
0.35 & 480 & 4.107(2) & 5.428(3) & 79.28(7) & 100 & 12.0 & 13.0\\
\end{tabular*}
\end{footnotesize}
\end{table*}

\begin{table*}
\caption{\label{tab:XRD}
Parameters derived from Rietveld refinement of the XRD patterns: temperatures of the structural phase transition
on heating $T^{(h)}_{s}$ and cooling $T^{(c)}_{s}$ together with phase fractions found at low temperatures (80-100~K)
and 480~K.}
\begin{footnotesize}
\begin{tabular*}{\textwidth}{@{\extracolsep{\fill}}cccccc}
$x$ & 0.05 & 0.10 & 0.15 & 0.22 & 0.35 \\ \hline
temperature range [K] & 100-480 & 83-480 & 83-480 & 80-480 & 80-480 \\
$T^{(h)}_{s}$ & 440 & 286 & 167 & - & - \\
$T^{(c)}_{s}$ & 393 & 245 & 140 & - & - \\
weight fraction of & 11.8(8) & 3.6(5) & 17.8(7) & 100$^*$ & 100$^*$ \\
hexagonal phase at 80-100~K [\%] & & & & \\
weight fraction of & 83.2(3.7) & 100 & 100 & 100$^*$ & 100$^*$ \\
hexagonal phase at 480~K [\%] & & & & \\ \hline
\end{tabular*}
\vspace{-.4cm}
\begin{flushleft}\noindent $^*$Small impurity peaks have been observed but they have been excluded from fitting,
as they are too weak to be identified unambiguously.\end{flushleft}
\end{footnotesize}
\end{table*}

Rietveld refinement of the XRD patterns has allowed to determine temperature evolution of the structural
parameters as well as the phase fraction (see Fig.~\ref{fig:temp_depend_of_cryst_param_x=0.05-0.15} and Tables~\ref{tab:latticeort}, \ref{tab:latticehex} and~\ref{tab:XRD}).
The TiNiSi-type crystal structure, which is dominant at low temperatures, transforms into the hexagonal
Ni$_2$In-type structure with increasing temperature. The transformation process is characterized by the presence
of a distinct hysteresis (see the upper insets in Fig.~\ref{fig:temp_depend_of_cryst_param_x=0.05-0.15}), characteristic
of the first-order phase transition. The critical temperature of the phase transition decreases
with increasing Cu content.

Considering the relationship $a_o=c_h$, $b_o=a_h$, $c_o=\sqrt{3}a_h$ and $V_o=2V_h$ (where the ``o'' and ``h''
indices refer to the orthorhombic and hexagonal structures, respectively), it is found that the martensitic
transformation is accompanied by a large volume expansion $\frac{V_o-2V_h}{V_o}$ of about 4~\%, as well as
strains of the lattice $\frac{a_o-a_h}{a_o}$ and $\frac{b_o-b_h}{b_o}$ equal to about 11~\% and 7~\%,
respectively. Such results indicate large entropy change associated with the martensitic transformation.
In case the magnetocaloric coupling is present, a large entropy magnetocaloric effect is expected to
accompany the transformation.

\begin{figure}[!ht]
\begin{center}
\includegraphics[bb=14 14 581 496, width=\columnwidth]
	{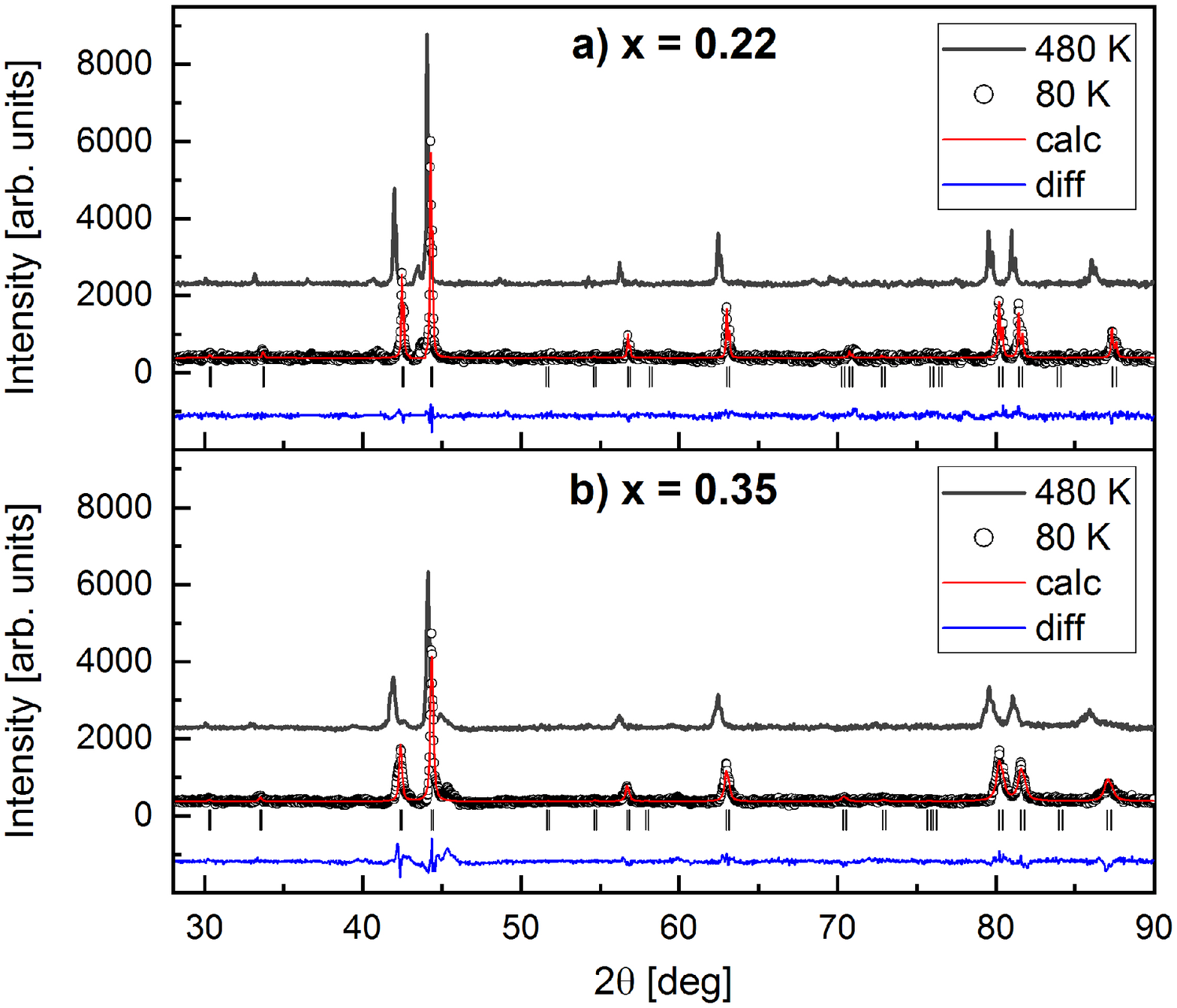}
\end{center}
\caption{\label{fig:XRD_patterns_x=0.22_0.35}
XRD patterns of Co$_{1-x}$Cu$_x$MnGe ($x=$0.22 and 0.35) collected at 480 and 80~K, together with the
results of Rietveld refinement of the patterns taken at 80~K.
For better clarity, the refined background has been subtracted from both the experimental
and calculated patterns after the final refinement.
The vertical bars indicate the Bragg reflection positions originating from the hexagonal phase
(space group $P6_3/mmc$).}
\end{figure}

\begin{figure}[!ht]
\begin{center}
\includegraphics[bb=14 14 865 610, width=\columnwidth]
	{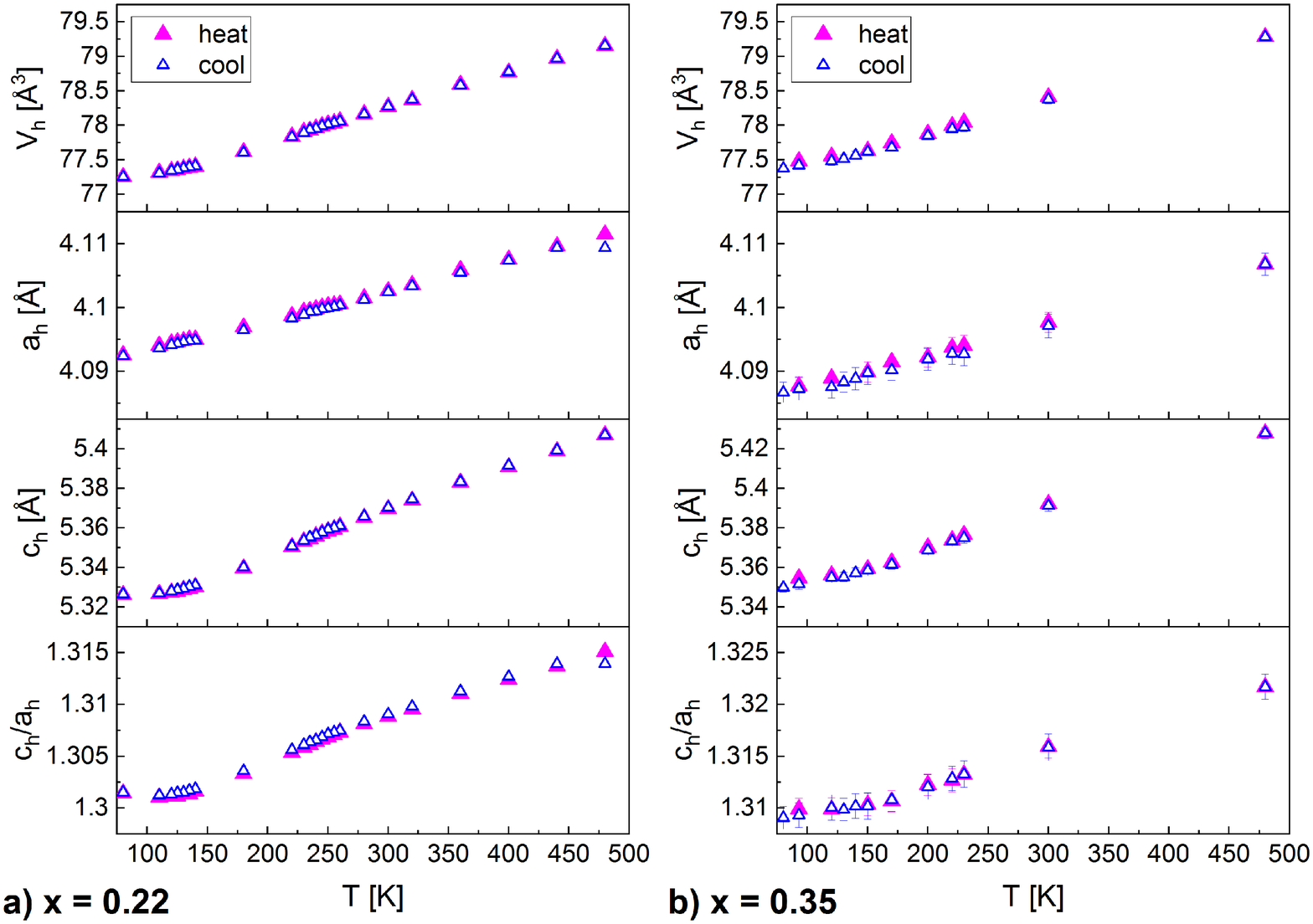}
\end{center}
\caption{\label{fig:temp_depend_of_cryst_param_x=0.22_0.35}
Thermal evolution of the volume of the hexagonal unit cell, lattice constants and their ratio
for Co$_{1-x}$Cu$_x$MnGe ($x=$0.22 and 0.35).}
\end{figure}

Fig.~\ref{fig:XRD_patterns_x=0.22_0.35} shows the XRD patterns of Co$_{1-x}$Cu$_x$MnGe ($x=$0.22 and 0.35)
collected at 80 and 480~K, together with the Rietveld refined data of the pattern taken at 80~K.
The experimental data indicate that the hexagonal crystal structure is stable in the whole temperature range.
The lattice parameters $a_h$ and $c_h$ as well as unit cell volume $V_h$ (see
Fig.~\ref{fig:temp_depend_of_cryst_param_x=0.22_0.35}) show nearly linear thermal expansion.

\subsection{Magnetic properties}

High pressure measurements have been carried out in the temperature range 80--400~K in fully hydrostatic
conditions. Helium has been used as a pressure-transmitting medium. The experimental cell has been connected
to the UNIPRESS OCA GCA-10 three stage gas compressor by a manganin gauge placed in the highest pressure
stage of the compressor. A thermocouple placed directly at the sample position has served as a temperature
sensor. The ac magnetic susceptibility measurements have been carried out in a weak magnetic field (of 1~mT
amplitude and frequency of 300~Hz). The voltage induced in the pick-up coils has been measured by a lock-in amplifier.

\begin{figure*}[!ht]
\begin{center}
\includegraphics[bb=14 14 635 524, height=0.33\textwidth]
	{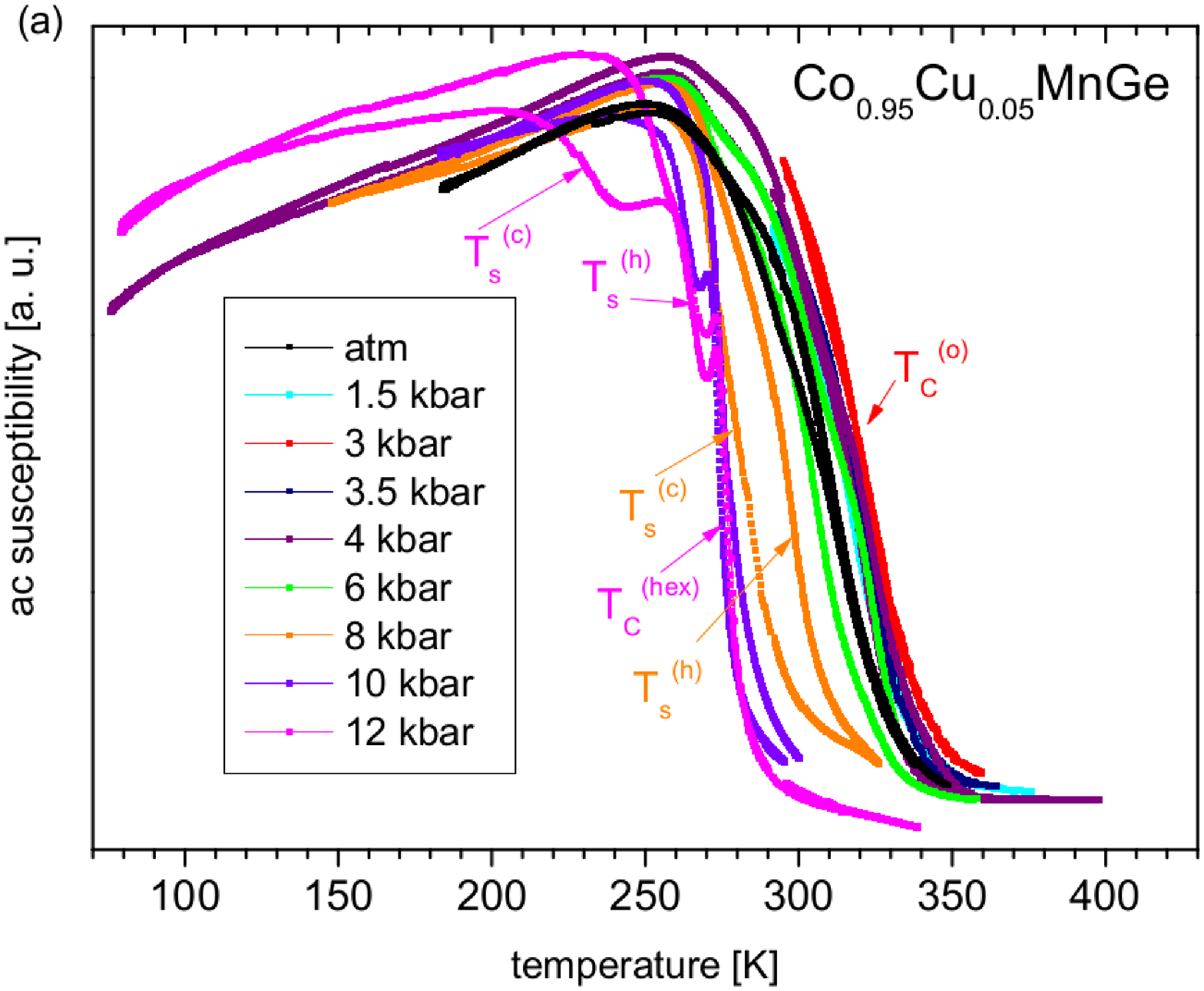}
\includegraphics[bb=14 14 637 523, height=0.33\textwidth]
	{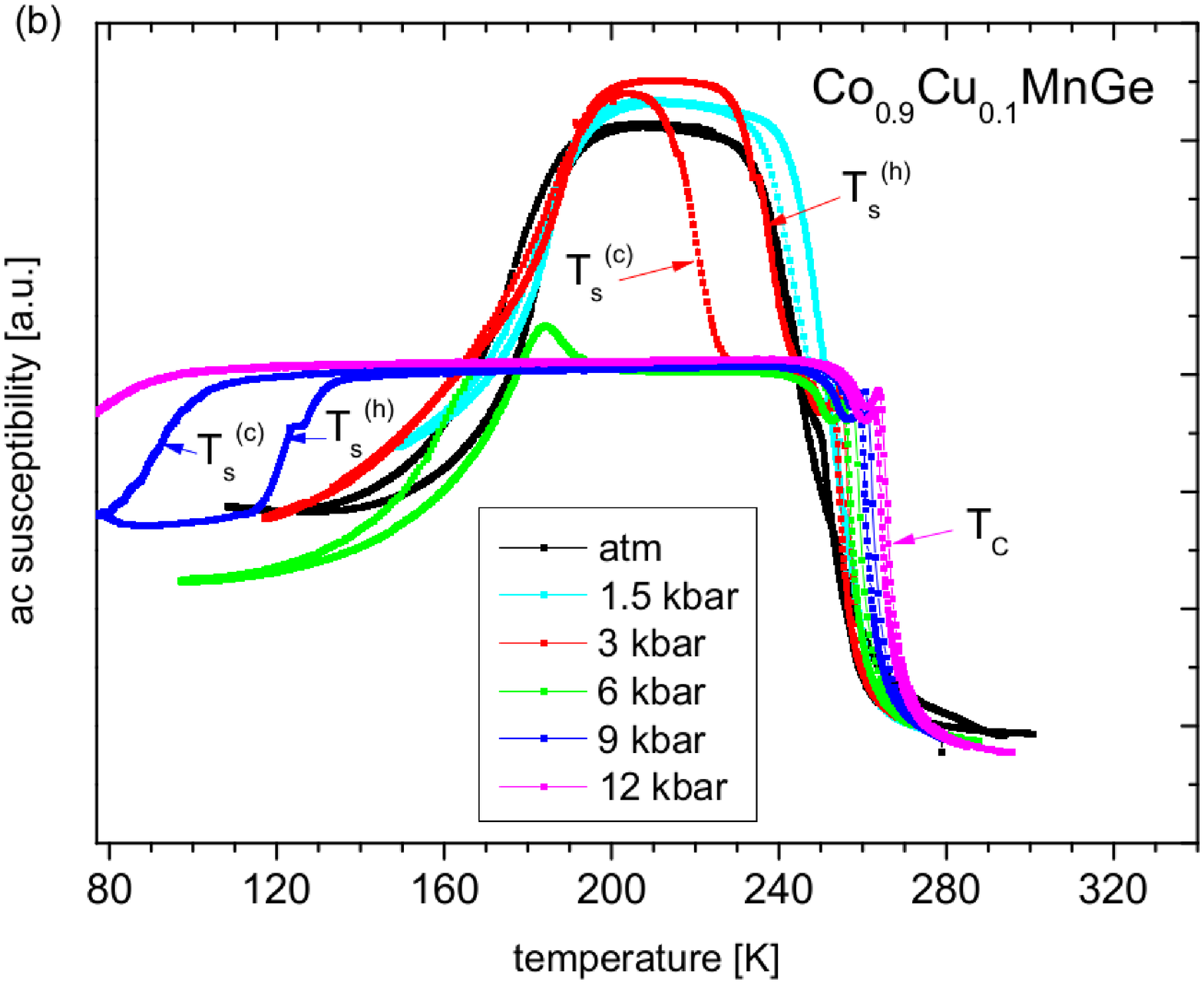}
\includegraphics[bb=14 14 650 517, height=0.33\textwidth]
	{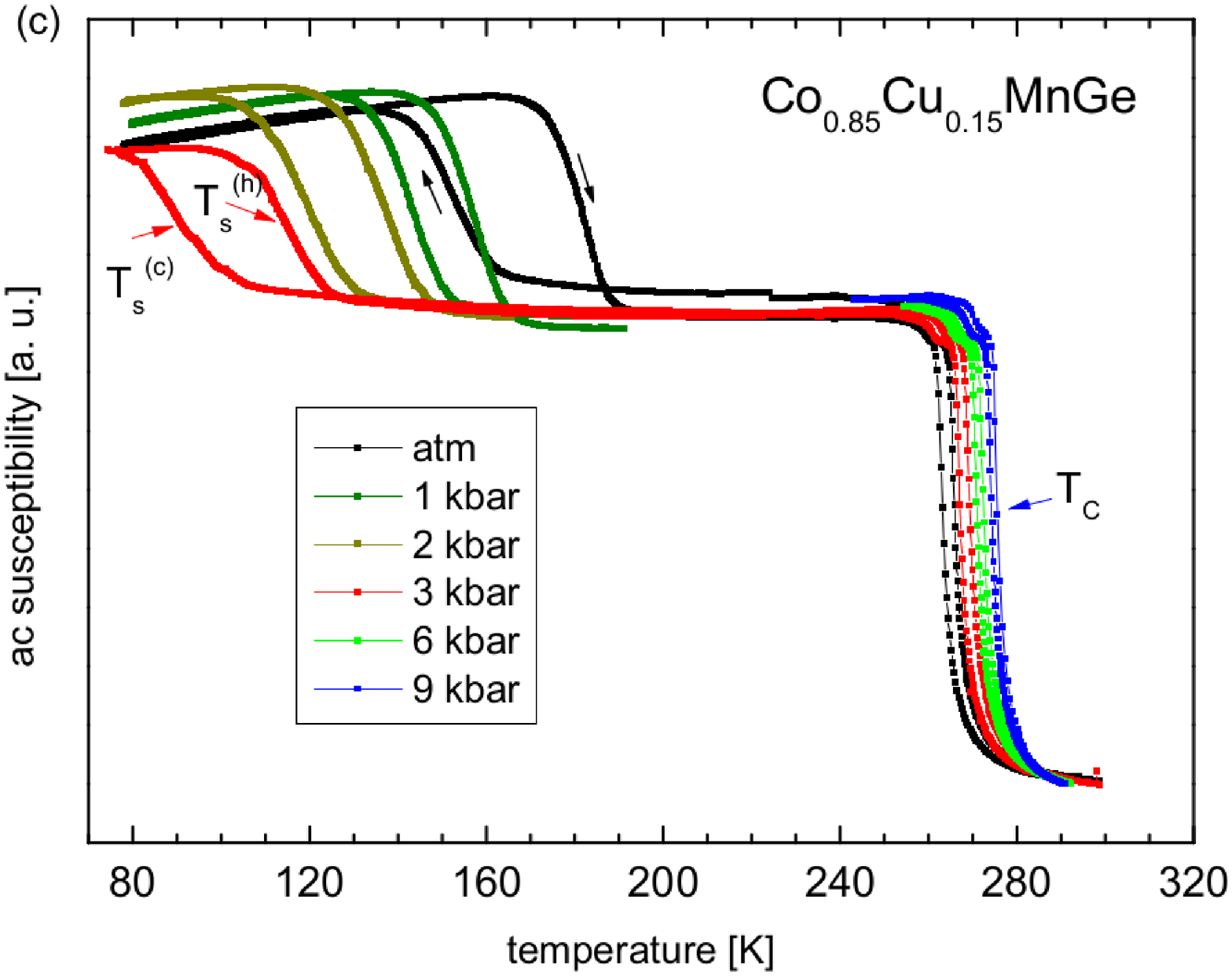}
\includegraphics[bb=14 14 637 519, height=0.33\textwidth]
	{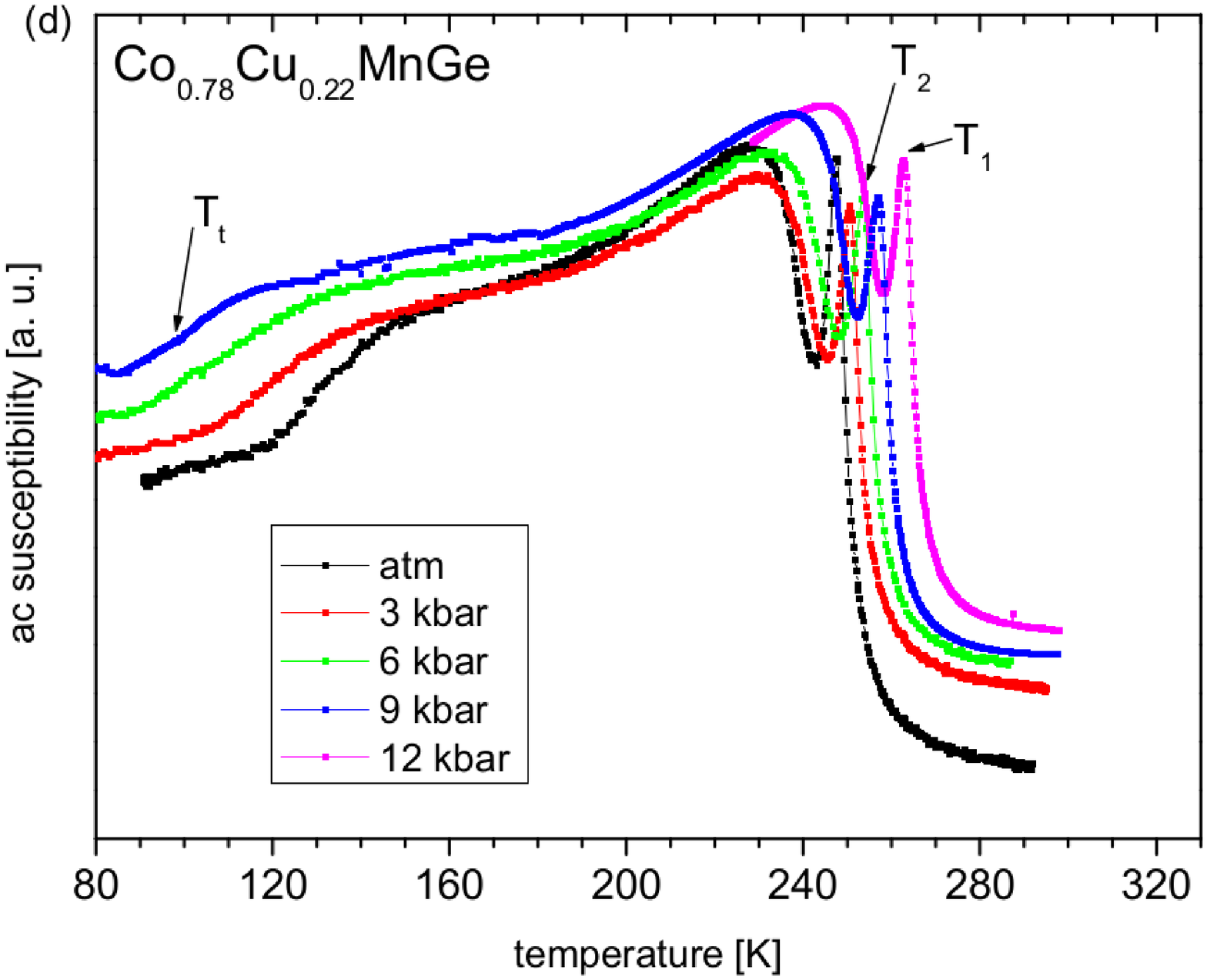}
\includegraphics[bb=14 14 660 522, height=0.33\textwidth]
	{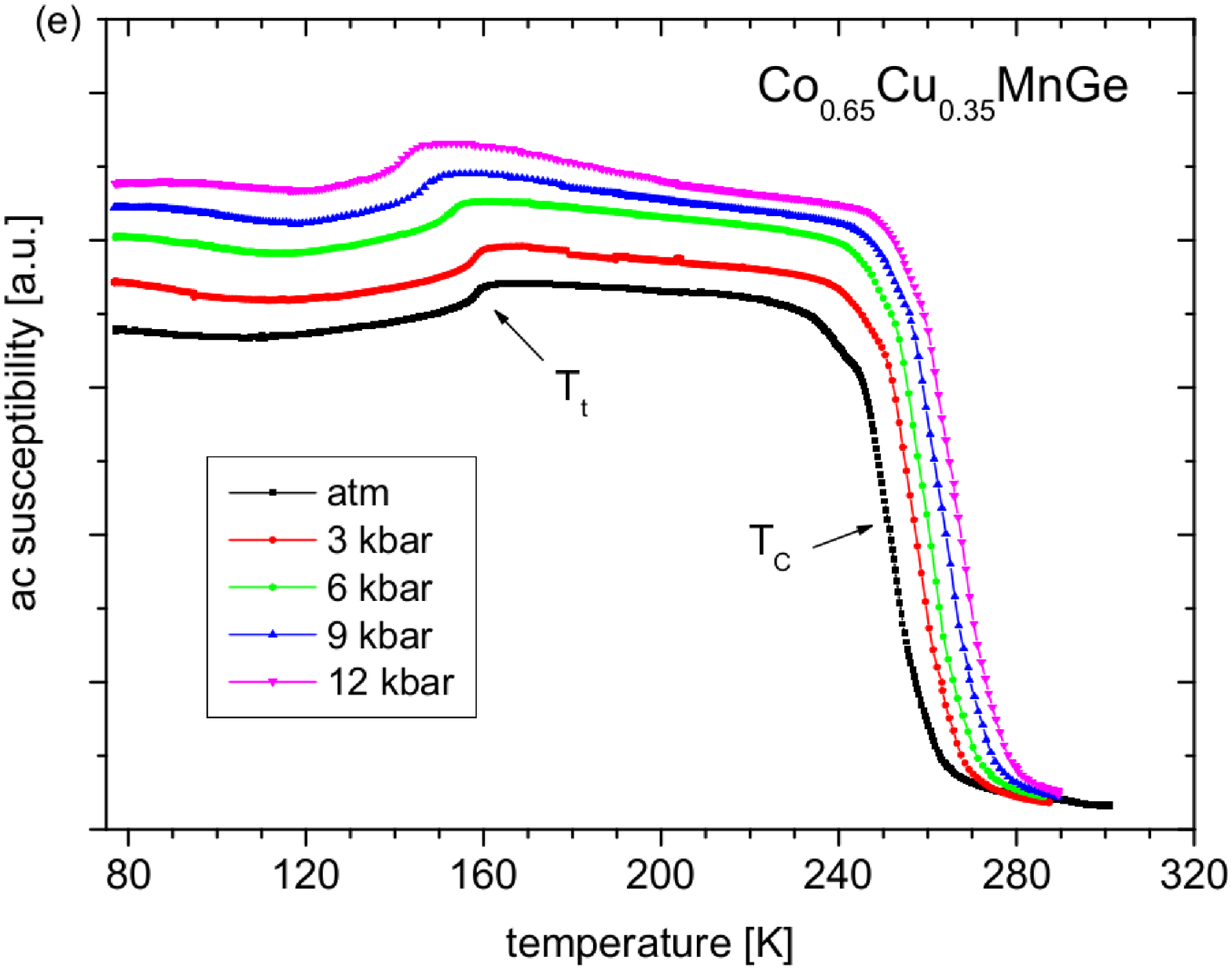}
\end{center}
\caption{\label{fig:ac_magn_susc}
Temperature dependence of the ac magnetic susceptibility of slowly cooled Co$_{1-x}$Cu$_x$MnGe for (a) $x=0.05$,
(b) $x=0.10$, (c) $x=0.15$, (d) $x=0.22$ and (e) $x=0.35$ at selected hydrostatic pressures up to 12~kbar.
}
\end{figure*}

Figs.~\ref{fig:ac_magn_susc}a--e show temperature dependence of the ac magnetic susceptibility collected
on heating and cooling for the samples with $x=0.05$, 0.10, 0.15, 0.22 and 0.35 at selected hydrostatic pressures
up to 12~kbar. For all compounds a sharp increase of susceptibility is observed at the ambient pressure below
315~K ($x=0.05$), 251~K (0.10), 260~K (0.15) and 250~K (0.35) (the transition temperatures have been defined
as temperatures of the inflection points in the $\chi(T)$ curves; in case of $x=0.22$, the transition has two-step
character and is discussed in details later in this section). Such a behavior is characteristic of
para- to ferromagnetic transition as observed in the hexagonal stoichiometric CoMnGe\cite{SZYTULA1981176}.
For the samples with $x \ge 0.10$ another anomalies appear with further decrease of temperature.
A decrease of magnetic susceptibility with hysteresis of a few kelvins is found around 180~K for $x=0.10$
indicating a development of antiferromagnetic component of the magnetic structure at low temperatures.
For $x=0.15$ an increase of susceptibility with wide temperature hysteresis ($\sim$30~K) is observed. This anomaly
well concides with the structural martensitic transition detected from the XRD data (compare
Figs.~\ref{fig:temp_depend_of_cryst_param_x=0.05-0.15}b and \ref{fig:ac_magn_susc}b).
The magnetic data for $x=0.22$ reveal complex magnetic properties -- with decrease of temperature a maximum at
$T_1=248$~K is followed by an inflection point at $T_2=238$~K and another inflection point at much lower temperature
$T_t=132$~K. Such a behavior suggests the following sequence of magnetic ordering: para-antiferro-ferro- and finally
ferrimagnetic phase. The sample with $x=0.35$ undergoes a para-ferromagnetic transition at $T_C=250$~K
followed by the transition to the final ferrimagnetic state at $T_t=159$~K.

Application of hydrostatic pressure has a significant influence on the magnetic properties, namely:

\begin{itemize}

\item $x=0.05$: For low pressures (up to 4~kbar), the Curie temperatures derived from the $\chi(T)$ curves
collected on cooling and heating agree within the accuracy of measurement. No thermal hysteresis indicates
a second-order phase transition. With increasing pressure ($p>6$~kbar) a distinct thermal hysteresis (up to 20~K),
characteristic of the first-order magnetostructural phase transition, develops. Moreover, for the highest pressures
(10 and 12~kbar), the transition from the paramagnetic state undergoes in two stages: with decreasing temperature
a typical second-order para-ferromagnetic transformation with no thermal hysteresis occurs in the hexagonal
Ni$_2$In-type phase, while further decrease of temperature leads to appearance of the first-order
magnetostructural phase transition associated with visible hysteresis (see Fig.~\ref{fig:ac_magn_susc}a).

\item $x=0.10$: A para-ferromagnetic transition shows a small hysteresis ($\sim$4~K) at ambient pressure indicating
partial magnetostructural coupling. With increasing pressure ($p=1.5$ and 3~kbar) this transition turns into two-step transition,
as observed previously for $x=0.05$ at high pressures (10 and 12~kbar), revealing temperature separation of
the purely magnetic and magnetostructural transitions. For $p\le 3$~kbar additional transition to
probably ferrimagnetic state appears at $T_t \approx 180$~K. A small hysteresis ($\sim$5~K) suggests a first-order
character of the latter transition. For higher pressures (6~kbar $\le p \le 9$~kbar) a purely magnetic
para-ferromagnetic transition (no temperature hysteresis visible) remains well-separated from the
ferro-ferrimagnetic magnetostructural transition (hysteresis of 10~K (6~kbar) or 25~K (9~kbar)). For the
highest pressure (12~kbar) only para-ferromagnetic transition is visible, while the magnetostructural transition
seems to be below the temperature range experimentally available (see Fig.~\ref{fig:ac_magn_susc}b).

\item $x=0.15$: A second order (no temperature hysteresis) para- to ferromagnetic transition undergoes at
$T_C = 264$~K, while the first order magnetostructural transition with distinct temperature hysteresis
is visible below 200~K. The temperature of the magnetostructural transition well coincides with martensitic
transition detected from the XRD data (compare Figs.~\ref{fig:temp_depend_of_cryst_param_x=0.05-0.15}c and \ref{fig:ac_magn_susc}c).

\item $x=0.22$ and 0.35: Application of external pressure influences transition temperatures, while the shape
of the $\chi(T)$ curves remains unchanged (see Figs~\ref{fig:ac_magn_susc}d and \ref{fig:ac_magn_susc}e).
All observed magnetic transitions show no detectable temperature hysteresis. For $x=0.22$, decrease of temperature
reveals a maximum in $\chi(T)$ at $T_1 = 248$~K followed by a rapid increase of magnetic susceptibility
at $T_2$ being about 10~K below $T_1$. The first anomaly suggests para- to antiferromagnetic transition at $T_1$
with further transformation to ferro- or ferrimagnetic state at $T_2$. Below 150~K another anomaly (inflection point)
is visible at $T_t$. The observed decrease of susceptibility suggests development of antiferromagnetic contribution
to the magnetic structure. For $x=0.35$, a rapid increase of susceptibility at $T_C = 250$~K proves
existence of para- to ferromagnetic transition, followed by a second transition (visible as inflection point
in $\chi(T)$) at $T_t = 159$~K -- the latter one evidencing for development of antiferromagnetic contribution
to the magnetic structure. It is worth noting that all the above mentioned magnetic transitions refer to the
hexagonal crystal structure which is found to exist within the investigated temperature interval at ambient pressure
(according to the XRD data) and no signs of structural transition are detected under applied hydrostatic pressure.

\end{itemize}

\begin{figure*}[!ht]
\begin{center}
\includegraphics[bb=14 14 676 524, height=0.33\textwidth]
	{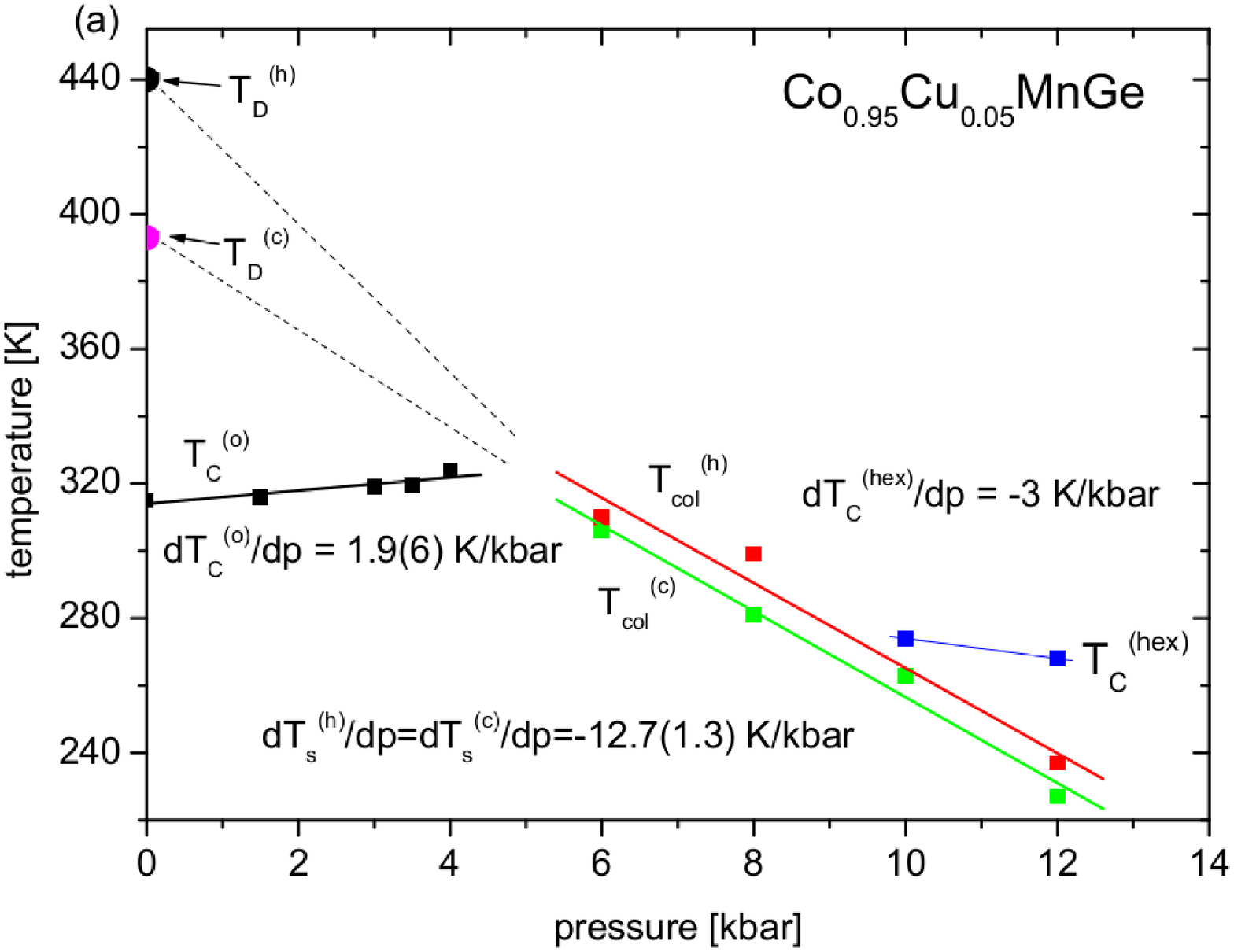}
\includegraphics[bb=14 14 667 521, height=0.33\textwidth]
	{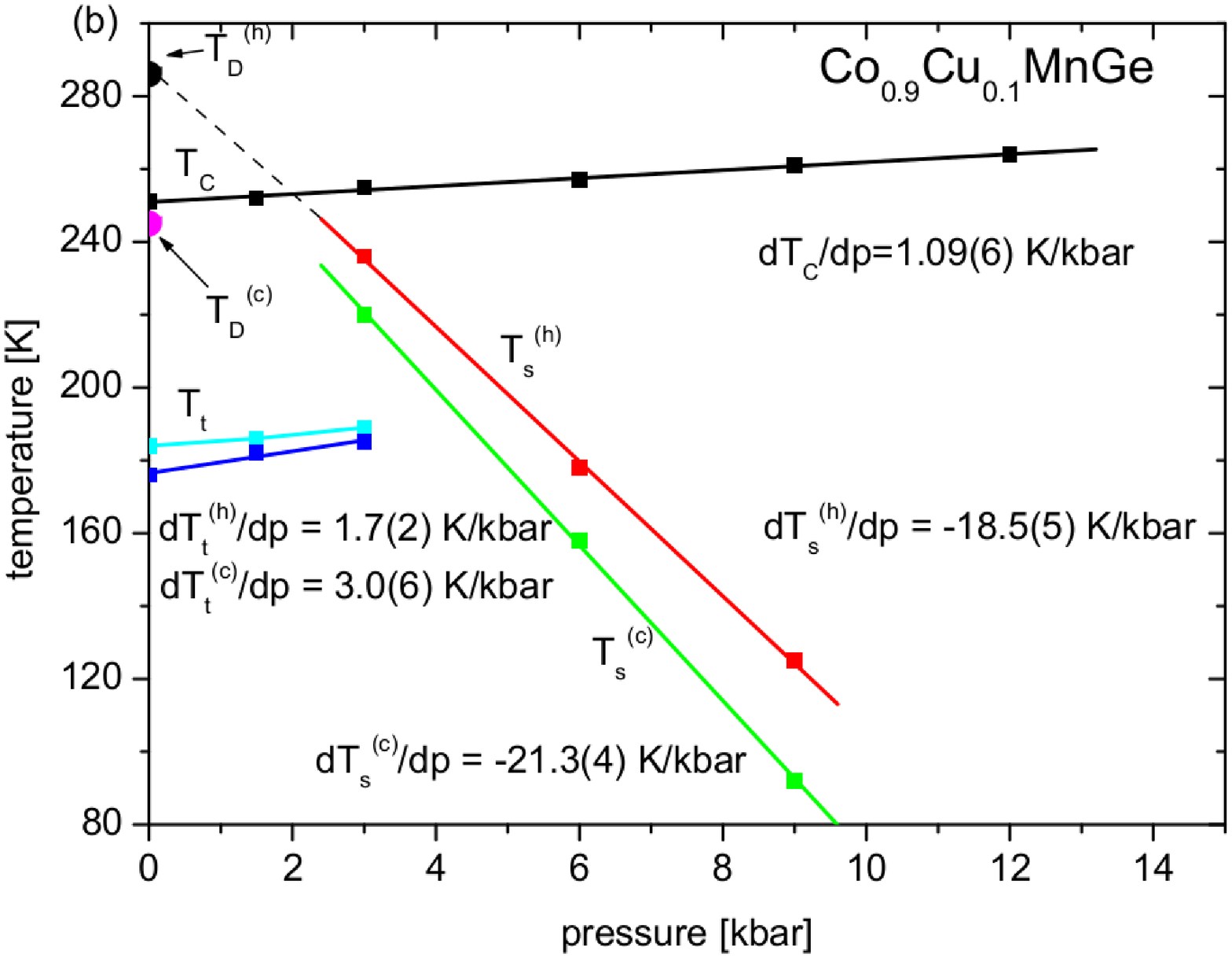}
\includegraphics[bb=14 14 672 514, height=0.33\textwidth]
	{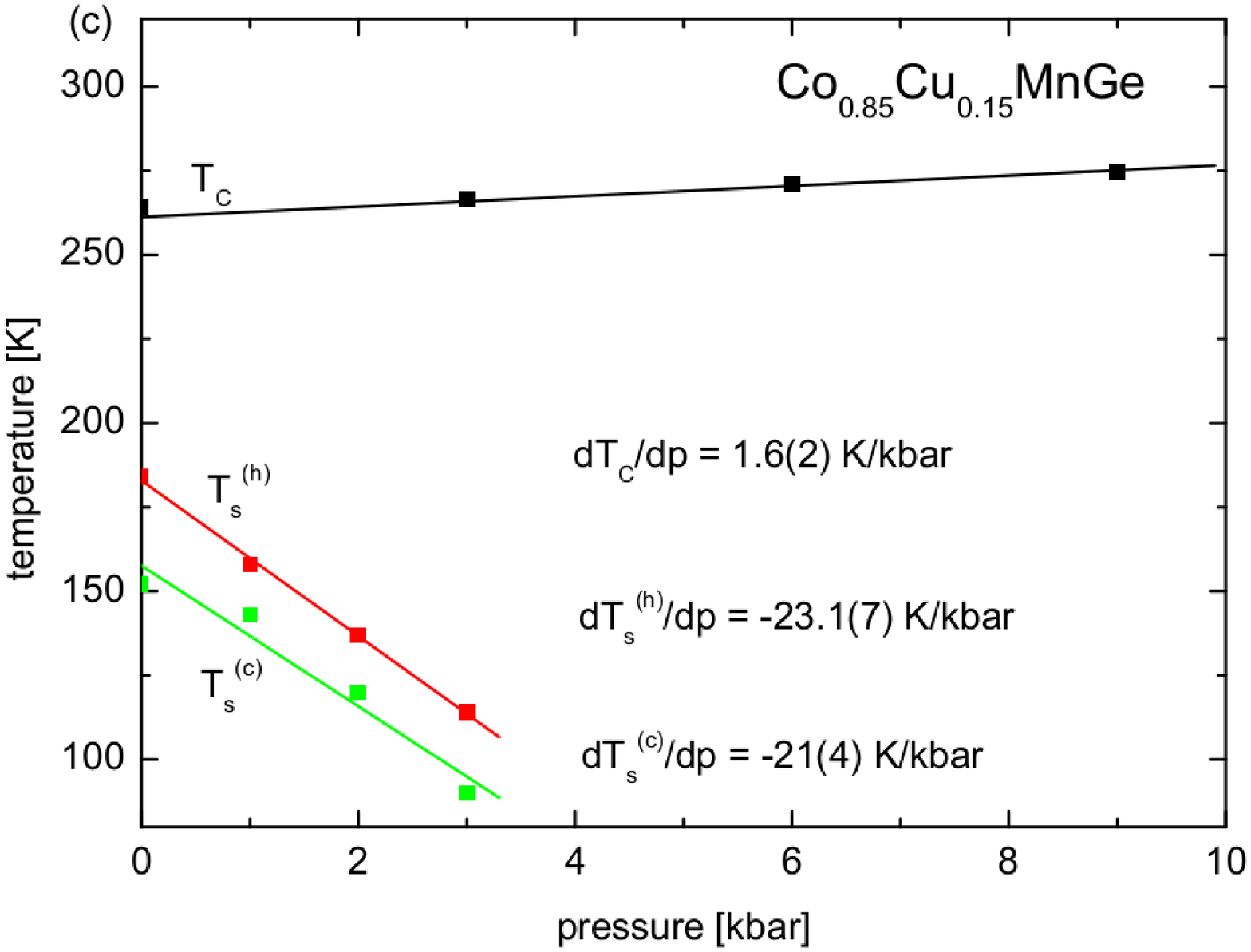}
\includegraphics[bb=14 14 670 523, height=0.33\textwidth]
	{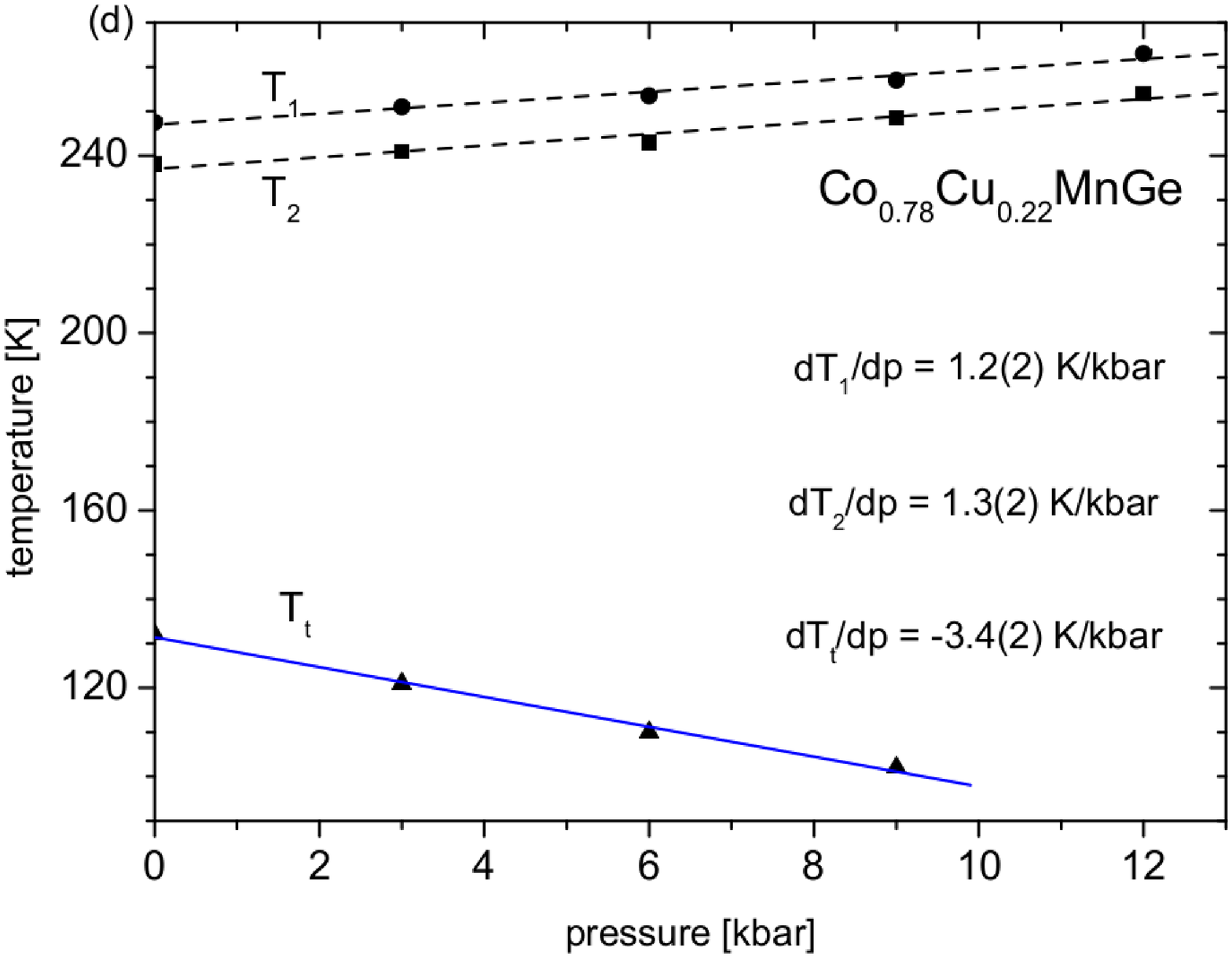}
\includegraphics[bb=14 14 681 531, height=0.33\textwidth]
	{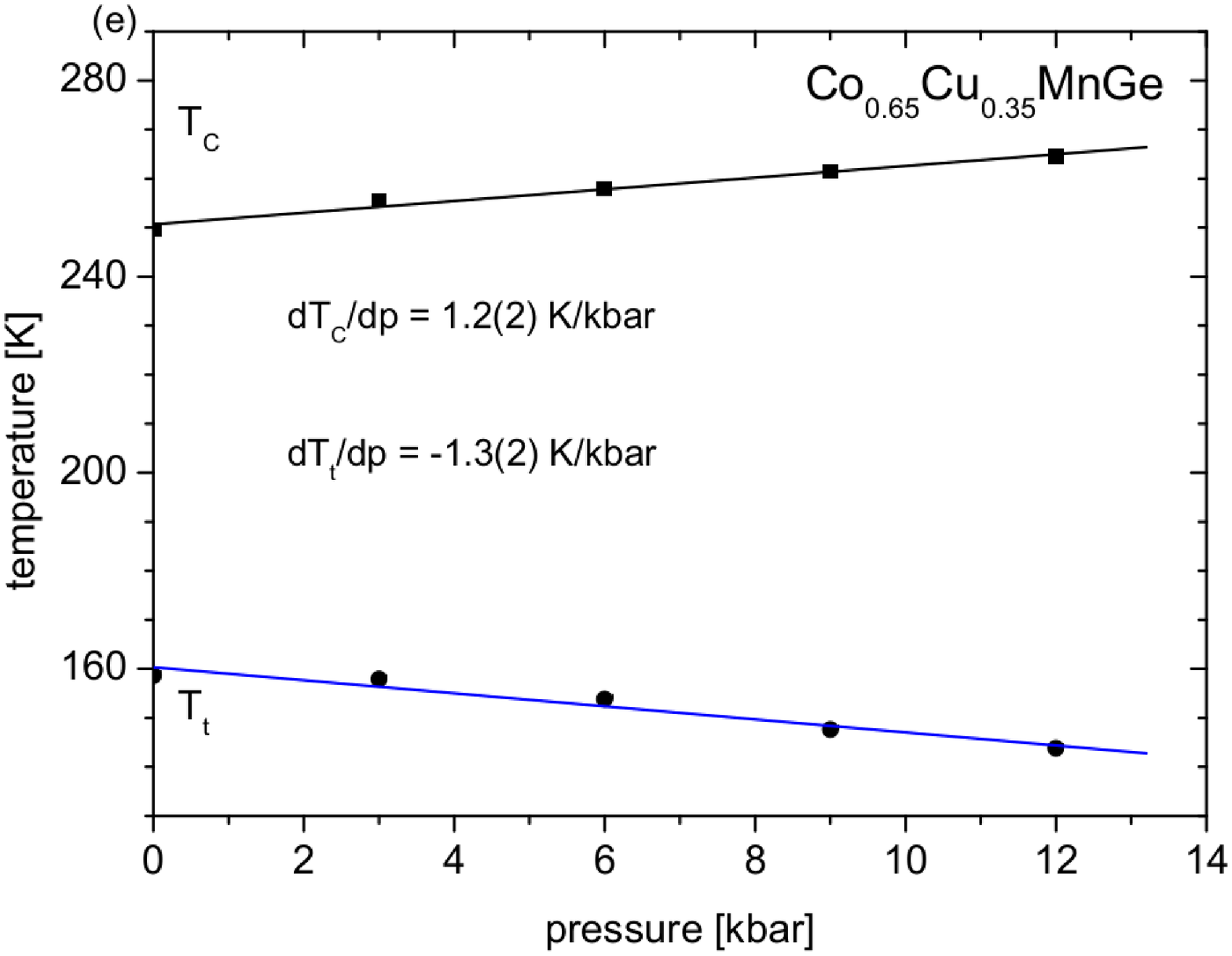}
\end{center}
\caption{\label{fig:p-T_diagram}
Magnetostructural (p,T) phase diagrams of slowly cooled Co$_{1-x}$Cu$_x$MnGe for (a) $x=0.05$, (b) $x=0.10$,
(c) $x=0.15$, (d) $x=0.22$ and (e) $x=0.35$. The (h) and (c) indexes refer to the heating and cooling processes,
respectively. $T_C$ and $T_s$ refer to the Curie and structural transition temperatures as derived from the
magnetometric data, respectively. The structural transition temperature determined from the XRD data
is denoted as $T_D$. The remaining magnetic transition temperatures are marked as $T_1$, $T_2$ and $T_t$
-- see the main text for details.
}
\end{figure*}

On the basis of the experimental data, the magnetostructural pressure-temperature $(p,T)$ phase diagrams have
been determined (see Figs.~\ref{fig:p-T_diagram}a--e). Although the magnetic properties are significantly
influenced by application of external pressure, some common features are visible in the phase diagrams:

\begin{itemize}

\item The temperature of structural martensitic transition $T_s$ is found to decrease almost linearly with increasing
hydrostatic pressure for $x = 0.05, 0.1$ and 0.15. The structural transition is fully ($x=0.05$; 6~kbar $\le p \le 8$~kbar)
or partially ($x=0.1$; ambient pressure) coupled with the transition from para- to ferromagnetic state for
selected chemical compositions and pressure ranges. No structural transition is detected for high Cu content ($x=0.22$
and 0.35).

\item The temperature of transition from para- to magnetically ordered state ($T_C$ for $x=0.1$, 0.15 and 0.35; $T_1$ for $x=0.22$)
increases linearly with increasing hydrostatic pressure. The $x=0.05$ case is more complicated, as $T_C$ increases
linearly with pressure up to $p=4$~kbar, then decreases linearly with pressure up to 8~kbar as the magnetic transition
is fully coupled with the structural one for intermediate pressures, and finally decreases linearly with lower rate
as application of high pressures ($p \ge 10$~kbar) leads to separation of $T_C = 275$~K and $T_s$ which is found below $T_C$.
It is worth noting that for $x=0.22$, the temperature of a second magnetic transition ($T_2$) depends on pressure
in the same way as $T_1$, i.e. both temperatures increase with increasing pressure with the same rate. (see Fig.~\ref{fig:p-T_diagram}d).

\item Additional magnetic transitions, occurring below $T_C$ and marked here as $T_t$, are observed for $x=0.1$, 0.22 and 0.35.
For $x=0.1$, $T_t \approx 180$~K appears only for lower pressures ($p \le 3$~kbar), where $T_t < T_s$. For $x=0.22$ and 0.35,
$T_t$ equals around 130~K ($x=0.22$) or 160~K ($x=0.35$) at ambient pressure and linearly decreases with applied
hydrostatic pressure.

\end{itemize}

\begin{table*}
\caption{\label{tab:magnetic}
Parameters characterizing pressure-temperature $(p,T)$ phase diagrams of slowly cooled Co$_{1-x}$Cu$_x$MnGe
($0 \le x \le 0.35$): Curie temperatures ($T_C$), other magnetic transition temperatures ($T_1, T_2, T_t$) and temperatures
of the structural transition ($T_s$), together with their derivatives over applied pressure ($dT_C/dp$, $dT_1/dp$, $dT_2/dp$,
$dT_t/dp$, $dT_s/dp$). The (h) and (c) indices refer to the heating and cooling processes, respectively.
$\gamma$ indicates the Gr\"uneisen parameter derived from the $\textrm{d}T_C/\textrm{d}p$ derivative -- see main text for details.
}
\begin{footnotesize}
\begin{tabular*}{\textwidth}{@{\extracolsep{\fill}}cccccccccc}
$x$ & & $T_C$ &$\textrm{d}T_C/\textrm{d}p$ & $T_t$ & & $\textrm{d}T_t/\textrm{d}p$ & $\textrm{d}T^{(h)}_{s}/\textrm{d}p$ &$\textrm{d}T^{(c)}_{s}/\textrm{d}p$ & $\gamma$\\
 & & [K] & [K/kbar] & [K] & & [K/kbar] & [K/kbar] & [K/kbar] \\ \hline
0$^*$ & & 340 & 3.6 & & & & -11.6 & -10.6 & 8.1\\
0.05 & & 313 & 1.9(6) & & & & -12.7(1.3) & -12.7(1.3) & 4.5\\
0.10 & & 250 & 1.09(6) & 180 & & 1.7(2) & -18.5(5) & -21.3(4) & 3.2\\
0.15 & & 260 & 1.6(2) & & & & -23.1(7) & -21(4) & 4.5\\
0.22 & & 247 ($T_1$) & 1.2(2) ($dT_1/dp$) & 135 & & -3.4(2) & & & 3.6\\
     & & 235 ($T_2$) & 1.3(2) ($dT_2/dp$) & & & & & & \\
0.35 & & 250 & 1.2(2) & 160 & & -1.3(2) & & & 3.6\\ \hline
\end{tabular*}
\end{footnotesize}
\begin{flushleft}$^*$ data from Ref.~\cite{NIZIOL1983205}.\end{flushleft}
\end{table*}

The parameters characterizing pressure-temperature $(p,T)$ phase diagrams of slowly cooled Co$_{1-x}$Cu$_x$MnGe
($0.05 \le x \le 0.35$) are summarized in Table~\ref{tab:magnetic}. For comparison, the data for stoichiometric CoMnGe
are also included.

\section{Discussion}

The work reports the results of X-ray diffraction as well as magnetic measurements under applied hydrostatic pressure
for Co$_{1-x}$Cu$_x$MnGe, where $x$ equals 0.05, 0.10, 0.15, 0.22 and 0.35.
The X-ray diffraction data
collected at low temperature confirm the orthorhombic structure for $x=0.05$, 0.10 and 0.15 and the hexagonal one
for the remaining compounds ($x=0.22$ and 0.35). With increase of temperature, the orthorhombic martensite structure
undergoes a martensitic transformation into the hexagonal austenite one. The transition temperature decreases
with increasing Cu content (see Table~\ref{tab:XRD}). The transition is associated with a distinct jump in the unit
cell volume $V_o$ and the lattice parameters $a_o$ and $b_o$ (see Figs.~\ref{fig:temp_depend_of_cryst_param_x=0.05-0.15}a--c).
Such a behavior has already been reported for the stoichiometric CoMnGe and is related to the changes in atom positional
parameters as well as to the Co thermal parameters being considerably larger in the high-temperature hexagonal
structure variant~\cite{Jeitschko:a12116}. It is worth noting that the atoms are more densely packed in the
high-temperature hexagonal Ni$_2$In-type structure when compared with the low-temperature orthorhombic TiNiSi-type one.

The results of magnetic measurements indicate a transition from para- to ferro-/ferrimagnetic state with decreasing
temperature (although the transition for $x=0.22$ involves an intermediate antiferromagnetic phase in limited
temperature range). The Curie temperature decreases from 313~K (in the orthorhombic phase) for $x=0.05$ to about
250~K (in the hexagonal phase) for the remaining compositions ($0.1 \le x \le 0.35$).

For all samples the pressure-temperature $(p,T)$ phase diagrams have been determined. For $x=0.05$, the Curie temperature
increases slowly with increasing pressure in the low-pressure range ($ p \le 4$~kbar) -- see Fig.~\ref{fig:p-T_diagram}a.
The derivative $\textrm{d}T_C/\textrm{d}p$ equals 1.9(6)~K/kbar and is lower than 3.6~K/kbar reported for the stoichiometric
\mbox{CoMnGe}~\cite{NIZIOL1983205}.
Further increase of pressure leads to appearance of a triple point $(p_{TRI},T_{TRI}) \approx (5\textrm{~kbar}, 325\textrm{~K})$, where the structural martensitic
transition coincides with para- to ferromagnetic transition. 
It is worth noting that both $p_{TRI}$ and $T_{TRI}$
take values lower than those found in the stoichiometric \mbox{CoMnGe} $\left( (p_{TRI},T_{TRI}) \approx (6\textrm{~kbar}, 360\textrm{~K}) \right) $~\cite{NIZIOL1983205}.
For pressures
$p_{TRI} \le p \le 8$~kbar, the para- to ferromagnetic transition remains fully coupled with martensitic
transition, which is confirmed by noticeable hysteresis of the susceptibility vs. temperature curves (see
Fig.~\ref{fig:ac_magn_susc}a). For the highest pressures $(10\textrm{~kbar} \le p \le 12\textrm{~kbar})$
the second-order para- to ferromagnetic transition (no visible hysteresis) becomes separated from the
martensitic transition (well visible hysteresis at lower temperatures). In this pressure range, the Curie temperature
decreases with increasing pressure with a rate of -3~K/kbar. The exact values of the $\textrm{d}T_C/\textrm{d}p$,
$\textrm{d}T^{(h)}_s/\textrm{d}p$ and $\textrm{d}T^{(c)}_s/\textrm{d}p$ derivatives,
as obtained on heating and cooling, are collected in Table~\ref{tab:magnetic}.
For comparison, the values for $x=0$ (CoMnGe) are also reported after Ref.~\cite{NIZIOL1983205}.

The $(p,T)$ phase diagrams for $x=0.1$ and 0.15 have common features: the Curie temperature ($T_C$) increases with
increasing pressure with a rate of 1.09(6)~K/kbar ($x=0.1$) or 1.6(2)~K/kbar ($x=0.15$), while the temperature of
structural transition ($T_s$) decreases with increasing pressure with a rate of about -20~K/kbar (the exact values
of the transition temperatures and their rates can be found in Table~\ref{tab:magnetic}. What distinguishes the
diagram for $x=0.1$ is partial magnetostructural coupling of the para- to ferromagnetic transition with the martensitic
one at ambient pressure, as well as presence of additional transition ($T_t \approx 180$~K) at low pressures ($0 \le p \le 3$~kbar).

The $(p,T)$ phase diagrams for $x=0.22$ and 0.35 are quite similar one to another as the temperature of transition
from para- to magnetically ordered state increases, while $T_t$ decreases with increasing pressure (see
Figs.~\ref{fig:p-T_diagram}d--e). The exact values of the transition temperatures as well as their derivatives
over pressure are listed in Table~\ref{tab:magnetic}.

It is worth noting that the derivatives $\textrm{d}T_C/\textrm{d}p$ in the hexagonal Ni$_2$In-type structure
($0.1 \le x \le 0.35$ -- see Figs.~\ref{fig:ac_magn_susc}b--e) equal about 1~K/kbar, which is close to
0.9(1)~K/kbar reported for CoMnGe in the metastable Ni$_2$In-type structure~\cite{https://doi.org/10.1002/pssa.2210840129}.

Magnetostructural phase diagrams reported in this work as well as those of other ternary and pseudoternary MM'Ge compounds
(where M and M' are transition elements) can be analyzed on the basis of the phenomenological Landau potential~\cite{Landau_Lifshitz_Statistical_Physics}:

\begin{equation}
\label{eq:Landau_potential}
\begin{aligned}
\Phi(\zeta,\eta,p,T) = {} & a[T-T_s(p)]\zeta^2 + e\zeta^4 + k\zeta^6\\
	& + b[T-T_C(p)]\eta^2 + f\eta^4 + g\zeta^2\eta^2
\end{aligned}
\end{equation}

\noindent

\noindent where $\zeta$ and $\eta$ refer to the structural and magnetic order parameters, respectively, while
$a$ (>0), $e$ (<0), $k,b,f$ (>0) and $g$ (<0) are phenomenological constants. The structural order parameter $\zeta$
describes a distortion of the crystal structure, while $\eta$ is related to magnetization. The terms in
Eq.~\ref{eq:Landau_potential} containing solely $\zeta$ or $\eta$ refer to the structural and magnetic transitions,
respectively, while the $g\zeta^2\eta^2$ term describes a magnetostructural coupling.

Minimizing the Landau potential given by Eq.~\ref{eq:Landau_potential} with respect to $\zeta$ and $\eta$ leads to
the following phases that may appear in the $(p,T)$ phase diagram: 

\begin{enumerate}[label=(\roman*)]

\item \label{PH} $\zeta=0$, $\eta=0$ \quad paramagnetic hexagonal phase,

\item \label{PO} $\zeta\neq 0$, $\eta=0$ \quad paramagnetic orthorhombic phase,

\item \label{FH} $\zeta=0$, $\eta\neq 0$ \quad ferromagnetic hexagonal phase,

\item \label{FO} $\zeta\neq 0$, $\eta\neq 0$ \quad ferromagnetic orthorhombic phase.

\end{enumerate}

On the $(p,T)$ phase diagram for $x=0.05$ (see Fig.~\ref{fig:p-T_diagram}a), the \ref{PH}, \ref{PO} and \ref{FO} regions are observed.
With increasing Cu content the~\ref{PO} region
vanishes, while the~\ref{FH} one develops (see Figs.~\ref{fig:p-T_diagram}b--c). For the highest Cu contents ($x=0.22$ and 0.35),
only the hexagonal Ni$_2$In-type crystal structure is observed, and therefore the \ref{PH} and \ref{FH} phases
are found in the corresponding phase diagrams (see Figs.~\ref{fig:p-T_diagram}d--e). The determined $(p,T)$ phase diagrams
for the investigated Co$_{1-x}$Cu$_x$MnGe compounds for $x=0.05$, 0.10 and 0.15 (see Figs.~\ref{fig:p-T_diagram}a--c)
are similar to these reported for the isostructural NiMn$_{1-x}$Cr$_x$Ge for $x=0.04$, 0.11. 0.18 and 0.25 (see Figs.~3a--d in
Ref.~\cite{DURAJ2018449}).

On the basis of the experimental data reported in this work, it is possible to calculate the entropy change connected with
the martensitic phase transition (both the forward (cooling) and reverse (heating) ones) directly from the Clausius-Clapeyron equation

\begin{equation}
\label{eq:Clausius-Clapeyron}
\Delta S = \frac{\Delta v}{\left( \frac{\textrm{d}T_s}{\textrm{d}p} \right) }
\end{equation}

\noindent where $\Delta v$ is a change of the unit cell volume per unit cell mass at the structural phase transition temperature $T_s$.
For $x=0.15$, $\Delta v = 5.2\cdot 10^{-6}$ m$^3 \cdot$ kg, according to the data shown in Fig.~\ref{fig:temp_depend_of_cryst_param_x=0.05-0.15}.
Taking into account the values of the $\textrm{d}T^{}_{s}/\textrm{d}p$ derivatives (see Fig.~\ref{fig:p-T_diagram}c), the entropy change connected
with the martensitic phase transition equals 22 and 25~J/(kg $\cdot$ K) for heating and cooling respectively.

Magnetic Gr\"uneisen parameter $\gamma$ is another quantity that can be calculated on the basis of the experimental data collected
in Table~\ref{tab:magnetic} using the relation

\begin{equation}
\label{eq:Gruneisen_parameter}
\gamma = \frac{\textrm{d} \ln T_C}{\textrm{d} \omega} = \frac{1}{\kappa T_C} \frac{\textrm{d} T_C}{\textrm{d} p}
\end{equation}

\noindent where $\omega$ is a relative volume change, while $\kappa$ is the compressibility (equal in this case
$1.36\cdot 10^{-3}$~kbar$^{-1}$ according to Ref.~\cite{Zach_PhD_thesis}). The Gr\"uneisen parameter can be
alternatively defined as $\frac{\textrm{d} \ln J}{\textrm{d} \omega}$, where $J$ is the effective exchange
integral. According to the data in Table~\ref{tab:magnetic}, substituting the Co atoms by the Cu atoms
gradually decreases the exchange integral.

In the hexagonal phase the Co atoms have no localized magnetic moments and therefore they do not participate in the
magnetic interactions~\cite{SZYTULA1981176}. The magnetic order is stable solely due to interactions between
the Mn magnetic moments. In the orthorhombic phase both the Mn and Co atoms carry magnetic moments~\cite{NIZIOL1982281} and the magnetic
order is stable due to interactions between the Mn-Mn, Mn-Co and Co-Co magnetic moments. Higher number of magnetic atoms
involved in the magnetic interactions lead to higher Curie temperatures observed in the orthorhombic phase.

The structural and magnetic properties of the investigated Co$_{1-x}$Cu$_x$MnGe system are strongly influenced by
two types of pressures: the chemical one (resulting from doping the Cu atoms) and the external applied one.
Substitution of the Co atoms by the Cu atoms induces positive chemical pressure as confirmed by gradual decrease
of the unit cell volume in the orthorhombic crystal phase with increasing Cu content (see Table~\ref{tab:latticeort} and Fig.~\ref{fig:temp_depend_of_cryst_param_x=0.05-0.15}).
Both types of pressure have similar influence on the crystal structure and magnetic properties, namely,
the increase of pressure stabilizes the hexagonal structure and leads to slow increase of the Curie temperature (except
the high pressure range ($p \ge 6$~kbar) for $x=0.05$, where full magnetostructural coupling is found for
6~kbar$ \ge x \ge 8$~kbar with further separation of the magnetic and structural transitions at higher pressures).

\section{Summary and conclusions}

Influence of temperature and applied hydrostatic pressure on structural and magnetic properties of the Co$_{1-x}$Cu$_x$MnGe
system ($x=0.05$, 0.10, 0.15, 0.22 and 0.35) is reported. Based on the experimental data the $(x,T)$ and $(p,T)$
magnetostructural phase diagrams have been determined. The presented results indicate that partial substitution of the Co
atoms by the Cu atoms strongly influences the physical properties of the investigated compounds, namely:

\begin{itemize}

\item a martensitic transition between the low-temperature orthorhombic crystal structure of the TiNiSi-type (space group:
$Pnma$) and the high-temperature hexagonal structure of the Ni$_2$In-type (space group: $P6_3/mmc$) is observed for
$x=0.05$, 0.10 and 0.15,

\item the martensitic transition from the orthorhombic to the hexagonal structure is accompanied with distinct decrease of
the unit cell volume $V$ and the lattice parameter $a$ as well as an increase of the lattice parameter $b$,

\item for the high Cu contents ($x=0.22$ and 0.35) only the hexagonal structure is observed,

\item doping the Cu atoms stabilizes the hexagonal crystal structure,

\item regardless the crystal structure, the compounds undergo a transition from para- to ferromagnetic state with
decreasing temperature (in case of $x=0.22$ through an intermediate antiferromagnetic phase). The para- to ferromagnetic transition is
fully coupled with the martensitic one for $x=0.05$ at the intermediate pressure range (6~kbar $\le p \le 8$~kbar).
Partial magnetostructural coupling is observed for $x=0.10$ at ambient pressure.

\item the Curie temperature at ambient pressure decreases from 313~K for $x=0.05$ (in the orthorhombic phase) to
about 250~K for the remaining compounds (in the hexagonal phase), 

\item temperature of the structural transition as well as the critical temperature of magnetic ordering decrease with
increasing Cu content,

\item application of hydrostatic pressure leads to decrease of temperature of the martensitic transition (for $x=0.05$,
0.10 and 0.15) and slow increase of the critical temperature of magnetic ordering (for all compositions, however, in case
of $x=0.05$ it refers to the low-pressure range ($p \le 4$~kbar) only),

\item the determined magnetostructural phase diagrams are explained based on the phenomenological Landau theory.

\end{itemize}

\section*{Acknowledgements}

We would like to thank Dr. Teresa Jaworska-Go\l{}\k{a}b from the M. Smoluchowski Institute of Physics,
Jagiellonian University, for discussions regarding the XRD measurements.

\end{document}